\documentclass[twocolumn,prl,aps,showpacs,superscriptaddress]{revtex4-2}
\usepackage{graphicx}
\usepackage{color}

\begin{document}

\title{Superconducting Proximity Effect in $(\sqrt{7}\times\sqrt{7})R19.1^{\text{o}}$ Ni Nanoislands on Pb(111)}

\author{Yen-Hui Lin}
\affiliation{Department of Physics, National Tsing Hua University, 30013 Hsinchu, Taiwan}
\author{Sucitto Teh}
\affiliation{Department of Physics, National Tsing Hua University, 30013 Hsinchu, Taiwan}
\author{Da-You Yeh}
\affiliation{Department of Physics, National Tsing Hua University, 30013 Hsinchu, Taiwan}
\author{Chin-Hsuan Chen}
\affiliation{Department of Physics, National Tsing Hua University, 30013 Hsinchu, Taiwan}
\author{Deng-Sung Lin}
\affiliation{Department of Physics, National Tsing Hua University, 30013 Hsinchu, Taiwan}
\affiliation{Center for Quantum Technology, National Tsing Hua University, Hsinchu 30013, Taiwan}
\author{Horng-Tay Jeng}
\email{jeng@phys.nthu.edu.tw}
\affiliation{Department of Physics, National Tsing Hua University, 30013 Hsinchu, Taiwan}
\affiliation{Center for Quantum Technology, National Tsing Hua University, Hsinchu 30013, Taiwan}
\affiliation{Physics Division, National Center for Theoretical Sciences, Hsinchu 30013, Taiwan}
\affiliation{Institute of Physics, Academia Sinica, Taipei 11529, Taiwan}
\author{Pin-Jui Hsu}
\email{pinjuihsu@phys.nthu.edu.tw}
\affiliation{Department of Physics, National Tsing Hua University, 30013 Hsinchu, Taiwan}
\affiliation{Center for Quantum Technology, National Tsing Hua University, Hsinchu 30013, Taiwan}

\date{\today}

\vspace{2cm}
\begin{abstract}

We have studied the proximity-induced superconductivity in $(\sqrt{7}\times\sqrt{7})R19.1^{\text{o}}$ Ni nanoislands by combing scanning tunnelling microscopy/spectroscopy (STM/STS) with density-functional theory (DFT) calculation. Through depositing Ni onto Pb(111) substrate at 80\,K, the monolayer Ni nanoislands with the $(\sqrt{7}\times\sqrt{7})R19.1^{\text{o}}$ surface structure have been fabricated, where the termination of Ni atoms at hexagonal close-packed (hcp) site is energetically preferred and the filling of 3\textit{d} orbitals from the charge transfer leads to the vanishing magnetic moment of Ni atoms. The topographic $(\sqrt{7}\times\sqrt{7})R19.1^{\text{o}}$ lattice as well as the asymmetric height contrast in atomic unit cell are further corroborated by the STM simulations. With high spatial and energy resolution, tunneling conductance ($\mathrm{d}I/\mathrm{d}U$) spectra have resolved an isotropic superconducting gap with $\Delta_{Ni(\sqrt{7}\times\sqrt{7})} \approx 1.29$ meV, which is slightly larger than $\Delta_{Pb} \approx 1.25$ meV. The temperature dependence of $\Delta_{Ni(\sqrt{7}\times\sqrt{7})}$ supports the substrate-induced superconducting proximity effect according to the same transition temperature $T_{c} \approx 7.14$ K with the Pb(111). The line spectroscopy has spatially mapped out the small increase of $\Delta_{Ni(\sqrt{7}\times\sqrt{7})}$, which could be explained by an enhanced electron-phonon interaction ($V_{ep}$) under the framework of Bardeen–Cooper–Schrieffer (BCS) theory as a manifestation of the hole doping of Pb(111) from the surface Ni atoms.



\pacs{68.37.Ef, 75.70.Ak, 71.15.Mb}

\end{abstract}

\maketitle

\section{INTRODUCTION}

When a normal metal has been brought into a direct contact with the conventional \textit{s}-wave superconductor, the proximity effect\cite{PGdeGennes,JJHauser,CWJBeenakker,CLambert} facilitates the penetration of Cooper pairs that permits superconductivity in normal metal, offering a promising approach to artificially stabilize superconductivity in a wide variety of low-dimensional materials. In particular, the coexistence of ferromagnetism and superconductivity\cite{AIBuzdin,FSBergeret} has been realized in the thin layered ferromagnet/superconductor systems where the superconducting correlations remain coherent within a certain finite distance. This achievement gives rise to the odd-frequency spin-triplet superconductivity and the singlet Cooper pairs close to the ferromagnet/superconductor interfaces can in turn become spin-polarized\cite{FSBergeret1,AKadigrobov,MEschrig}. The triplet supercurrents can thus be generated leading to the minimized dissipation of Joule heating and the long spin lifetimes in terms of spin transport applications\cite{RSKeizer,HYang,MEschrig1}.

Given the great advancements in making nanometer-sized magnetic materials superconducting via proximity effect, not only exceptional superconductivity, but also newly emergent quasiparticle states can be stabilized in the reduced dimensions. According to Kitaev's toy lattice model\cite{Kitaev}, one-dimensional (1D) topological superconductivity with a \textit{p}-wave-like pairing can host a novel type of quasiparticle excitation, i.e., Majorana fermions (MFs), which can be achieved by engineering the interplay between ferromagnetism and conventional superconductivity with strong spin-orbit interaction in condensed matter systems. By means of growing ferromagnetic Fe chains on top of Pb(110), Nadj-Perge \textit{et al.}\cite{SNadjPerge} have recently reported the observation of Majorana zero modes (MZMs) bound to the ends of Fe atom chains and established the spin polarization measurements to distinguish the MZMs from trivial in-gap states\cite{SJeon}. Besides the magnetic atom chains\cite{MRuby,HKim}, such theoretical scheme can be further extended to support the emergence of MZMs at the edges of magnetic nanoislands in the 2D $\textit{p}_{x}+\textit{ip}_{y}$ topological superconductivity\cite{APMorales,GCMenard}. In perspective of intriguing property of being their own antiparticles, the MFs obey non-Abelian statistics in the adiabatic braiding processes and could be served as the building blocks for topological qubits, having an important potential for realizing fault-tolerant quantum computation\cite{JAlicea,CWJBeenakker1,DAasen}


In this work, we have investigated the proximity-induced superconductivity in $(\sqrt{7}\times\sqrt{7})R19.1^{\text{o}}$ Ni nanoislands by exploiting scanning tunnelling microscopy/spectroscopy (STM/STS) together with density-functional theory (DFT) calculation. By means of the low temperature growth at 80\,K, single-atomic-layer Ni nanoislands with $(\sqrt{7}\times\sqrt{7})R19.1^{\text{o}}$ surface structure have been fabricated on the Pb(111), where one unit cell accommodates 3 Ni atoms terminated at hcp site for the lower adsorption energy. As a result of charge transfer between Ni and Pb atoms, there is either zero magnetic moment of the Ni atoms or the absence of long-range magnetic ordering in the $(\sqrt{7}\times\sqrt{7})R19.1^{\text{o}}$ Ni nanoislands. The STM simulations comparably reproduce not only the topographic $(\sqrt{7}\times\sqrt{7})R19.1^{\text{o}}$ lattice, but also the asymmetric height contrast in the atomic unit cell. In addition to $\Delta_{Pb} \approx 1.25$ meV, the same "U" shape of isotropic superconductivity but a slightly larger $\Delta_{Ni(\sqrt{7}\times\sqrt{7})} \approx 1.29$ meV has also been measured by the tunnelling spectra, which is independent from structural changes at different atomic sites. The identical  $T_{c} \approx 7.14$ K has been extracted from the temperature-dependent $\Delta_{Ni(\sqrt{7}\times\sqrt{7})}$ and $\Delta_{Pb}$, further supporting the proximitized superconductivity in atomic-thick $(\sqrt{7}\times\sqrt{7})R19.1^{\text{o}}$ Ni nanoislands driven by the Pb(111) substrate. According to the hole doping of Pb(111) from the surface Ni atoms, an enhanced electron-phonon interaction ($V_{ep}$) on the basis of Bardeen–Cooper–Schrieffer (BCS) theory could offer an explanation for the small increase of $\Delta_{Ni(\sqrt{7}\times\sqrt{7})}$ as visualized directly in the spatial mapping from the line spectroscopy.

\section{EXPERIMENTAL AND THEORETICAL METHODS}

\textit{Experimental details.} The Ni/Pb(111) were prepared in an ultrahigh vacuum (UHV) chamber with the base pressure below $p \leq 2 \times 10^{-10}$\,mbar. The clean Pb(111) surface was first prepared by cycles of Ar$^{+}$ ion sputtering with an ion energy of 500\,eV at room temperature and subsequent annealing up to 600\,K. The Ni source with purity of 99.999\,\% (Goodfellow) was e-beam sublimated onto Pb(111) surface at the low temperature of 80\,K at which the well-ordered and uniform Ni nanoislands with $(\sqrt{7}\times\sqrt{7})R19.1^{\text{o}}$ lattice structures can be grown. 
After preparation, the sample was immediately transferred into a ultralow-temperature scanning tunneling microscope (LT-STM) from Unisoku Co. Ltd. (operation temperature $T \approx 0.32$\,K). The topography images were obtained from the constant-current mode with the bias voltage $U$ applied to the sample. For scanning tunneling spectroscopy (STS) measurements, a small bias voltage modulation was added to $U$ (frequency $\nu = 3991$\,Hz), such that tunneling differential conductance $\mathrm{d}I/\mathrm{d}U$ spectra can be acquired by detecting the first harmonic signal by means of a lock-in amplifier.

\textit{Theoretical computations.}

First-principles calculations are performed using Vienna Ab Initio Simulation Package (VASP)~\cite{GKresse} with local density approximation for the exchange-correlation functional~\cite{DMCeperley} based on density functional theory (DFT).
To simulate the experimental observations, we construct a structure model with $(\sqrt{7}\times\sqrt{7})R19.1^{\text{o}}$ Ni lattice on top of the 7-layer Pb(111) slab at the hcp sites using experimental lattice constant of 9.26~\AA~ and a vacuum thickness of 17.5~\AA. The structural optimization was carried out using the cutoff energy of 350~eV for the plane wave basis over the $6\times6\times1$ Monkhorst-Pack k-point mesh. The atomic positions of Ni atoms and top layer Pb(111) were relaxed until all the atomic forces were less than $10^{-3}$~eV/\AA. Formation energy calculations of Ni trimer model as well as other possible lattice models show that the current $(\sqrt{7}\times\sqrt{7})R19.1^{\text{o}}$ Ni lattice model is energetically the most favorable one.

The lattice dynamics calculations are performed on $48\times48\times48$ Monkhorst-Pack k-mesh with the cutoff energy of 60~Ry (600~Ry) for wavefunction (charge density) of Pb using density functional perturbation theory (DFPT) as implemented in the Quantum-Espresso code~\cite{PGiannozzi}. The electron-phonon coupling coefficient $\lambda_{qv}$ of mode $v$ and
wavevector $q$, is calculated with the interpolation over the Brillouin Zone using $4\times4\times4$ q-mesh.    
The effective Coulomb repulsion $\mu^\ast$ of 0.1~\cite{RDynes} is adopted. As the first step toward understanding the proximitized superconductivity induced by the Ni atoms on Pb(111), we calculated the superconducting temperature of bulk Pb as a function of hole-doping concentration, which corresponds to the charge transfer from Pb to Ni as discovered in our Bader charge analysis on the Ni trimer lattice model, using the Allen-Dynes modified McMillan formula~\cite{PBAllen},
 \[ T_C = f_1 f_2 \frac{\omega_{ln}}{1.2} exp (\frac{1.04 ( 1+ \lambda )}{\mu^\ast (1 + 0.62 \lambda)})\quad(1).\]
The detailed calculation of $f_1$ and $f_2$ can be found in Ref.~\cite{PBAllen}, while the logarithm average phonon frequency $\omega_{ln}$ and dimensionless coupling parameter $\lambda$ can be related to Eliashberg function $\alpha^2 F(\omega)$ as follows~\cite{PBAllen}:
\[\omega_{ln}=exp\left( \frac{2}{\lambda} \int \alpha^2 F(\omega) \frac{ln\omega}{\omega} d\omega \right), \quad(2)\]
\[\lambda=\sum_{qv} \lambda_{qv}=2\int \frac{\ \alpha^2 F(\omega)}{\omega} d\omega. \quad(3)\]


\section{RESULTS AND DISCUSSION}

\begin{figure}[htb]
\includegraphics[width=\columnwidth]{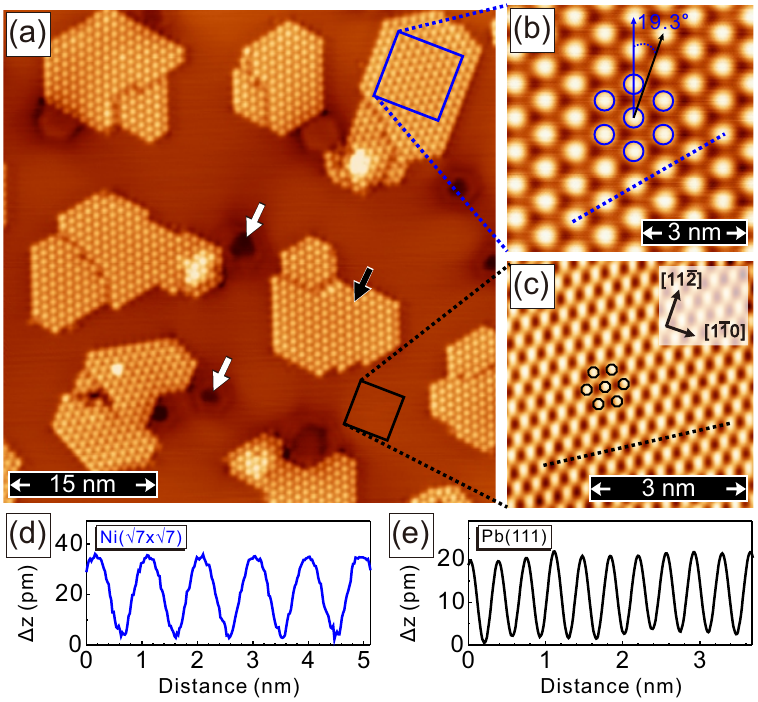}
\caption{(Color online)
(a)\, Overview of STM topography of Ni nanoislands with a surface coverage of about 0.4 ML after depositing Ni onto Pb(111) surface at 80 K. Two white arrows indicate surface features of Ar-induced nano-cavities after sputtering and annealing process on Pb(111) ($U_{b} = +1.0$\,V, $I_{t} = 0.4$\,nA). (b)\, Zoom-in image measured from the blue square in (a) shows the hexagonal $(\sqrt{7}\times\sqrt{7})R19.1^{\text{o}}$ lattice of Ni nanoislands, which can be derived from a direct comparison with the $(1\times1)$ atomic lattice of Pb(111) as shown in (c) ($U_{b} = +10$\,mV, $I_{t} = 1.0$\,nA). The line profiles in (d) and (e) correspond to the lattice constants of 9.57 \AA\, in (b) (blue dashed line) and 3.62 \AA\, in (c) (black dashed line), respectivley ($U_{b} = +10$\,mV, $I_{t} = 1.0$\,nA).
\label{fig1} }
\end{figure}

Fig. 1(a) is the STM topography overview of as-prepared sample after depositing Ni onto Pb(111) surface at low temperature 80\,K. The Ni nanoislands appear with a hexagonal lattice and have a surface coverage of about 0.4 ML. Two white arrows in Fig. 1(a) indicates typical surface features of nano-cavities formed by embedded Ar-ions after sputtering and annealing process on Pb(111)\cite{MSchmid,MMuller}. Fig. 1(b) displays the zoom-in image taken from the blue square on top of one Ni nanoisland in Fig. 1(a), where the hexagonal lattice represents a $19.3^{\text{o}}$ rotation with respect to the high symmetry crystalline direction of [11-2] determined by the $(1\times1)$ atomic lattice of Pb(111) as shown in Fig. 1(c). Note that if the hexagonal lattice of Ni nanoisland in Fig. 1(b) has a $19.3^{\text{o}}$ rotation, then the other Ni nanoisland, for example, the one indicated by black arrow in Fig. 1(a) corresponds to a $-19.3^{\text{o}}$ rotation, and these are the two different domains of Ni nanoislands that have been observed on the Pb(111) surface. Due to a threefold rotation symmetry of hexagonal lattice, each domain of Ni nanoislands should exhibit three rotational domains and they appear identical to each other. The lattice constants of of 9.57 \AA\, in Fig. 1(b) (blue dashed line) and 3.62 \AA\, in Fig. 1(c) (black dashed line) have been extracted from the corresponding line profiles in Fig. 1(d) and (c), respectively. Taking all these understandings into account, we are able to deduce that the Ni nanoislands with the $(\sqrt{7}\times\sqrt{7})R19.1^{\text{o}}$ hexagonal lattice based on the simple hard-sphere arguments\cite{MJHarrison,DFLi} have been grown on the Pb(111).

Between the two connected Ni nanoislands, there are either antiphase domain boundary or missing atom vacancies observed, which have been marked by red and green arrows in Fig. 2(a). Interestingly, these $(\sqrt{7}\times\sqrt{7})R19.1^{\text{o}}$ Ni nanoislands grown on the Pb(111) exhibit a small apparent height. For example, the line profile in Fig. 2(b) taken from the black dashed line in Fig. 2(a) shows a step height about 101 pm\, that is smaller than a single step height of either 286 pm\, on Pb(111) or 203 pm\, on Ni(111). In addition, we have also examined the bias dependence of apparent height and confirmed an average value of 88 pm\, for this small apparent height in the bias range of $\pm$ 1.0 V\, as the results shown in Fig. 2(c). In order to gain insights on the $(\sqrt{7}\times\sqrt{7})R19.1^{\text{o}}$ hexagonal lattice and unexpectedly small apparent height of Ni nanoislands, several attempts have been made to find the stable structure in the first-principles calculations and the results have been shown in Fig. 2(d). 

From the top view of structure model in Fig. 2(d), the Ni atoms prefer to stabilize at the hcp site on the Pb(111) surface, and each unit cell as marked by blue rhombus contains three Ni atoms in the $(\sqrt{7}\times\sqrt{7})R19.1^{\text{o}}$ hexagonal lattice. Although the same configuration of Ni atoms can be applied to the fcc-site termination, it has a higher energy cost, i.e., about 7 meV\, per atom, than the hcp-site termination, meaning more energetically unstable in terms of atomic adsorption kinetics. Apart from that, at the bottom panel of Fig. 2(d), the side view of relaxed structure model indicates that the Ni atoms tend to sink into the Pb(111) substrate and one Pb atom at the first surface layer surrounded by Ni atoms has been pushed a bit downward. After fully structural relaxation, the interlayer distance between surface Ni and Pb atoms turns out to be about 115 pm\,, which is supportive for the low apparent height observed experimentally. Moreover, based on the imaging mechanism of tunneling theory in STM\cite{JTersoff,EStoll,CJChen}, the contour of apparent height as a function of surface charge density has been calculated in Fig. 2(e). By comparing the $(\sqrt{7}\times\sqrt{7})R19.1^{\text{o}}$ Ni on Pb(111) (left) and the pristine Pb(111) surface (right), the apparent height difference is about 102 pm\, when the tip is about 6.3 \AA\, away from the surface. We denote that the apparent height further decreases to about 93 pm\, if tip-sample distance increases to 7.3 \AA\,, which remains comparable to the normal working distance about 10 \AA\, in the STM tunnelng junction\cite{JTersoff,EStoll,CJChen}.

To justify the validity of structure model in Fig. 2(d), the STM simulations have been further performed and the corresponding simulated STM image in Fig. 2(f) agrees well with the experimental STM topography as shown in Fig. 2(g). Note that the blue rhombus together with black circles refer to the unit cell of $(\sqrt{7}\times\sqrt{7})R19.1^{\text{o}}$ hexagonal lattice. More importantly, there is an asymmetric height contrast inside the unit cell as indicated by black and yellow arrows in Fig. 2(g), and this detailed feature has been consistently reproduced in the simulated STM image as well in Fig. 2(f), which is originated from the contribution of the Pb atoms of the second surface layer as indicated by the yellow arrow in the structure model of Fig. 2(d).


\begin{figure}[htb]
\includegraphics[width=\columnwidth]{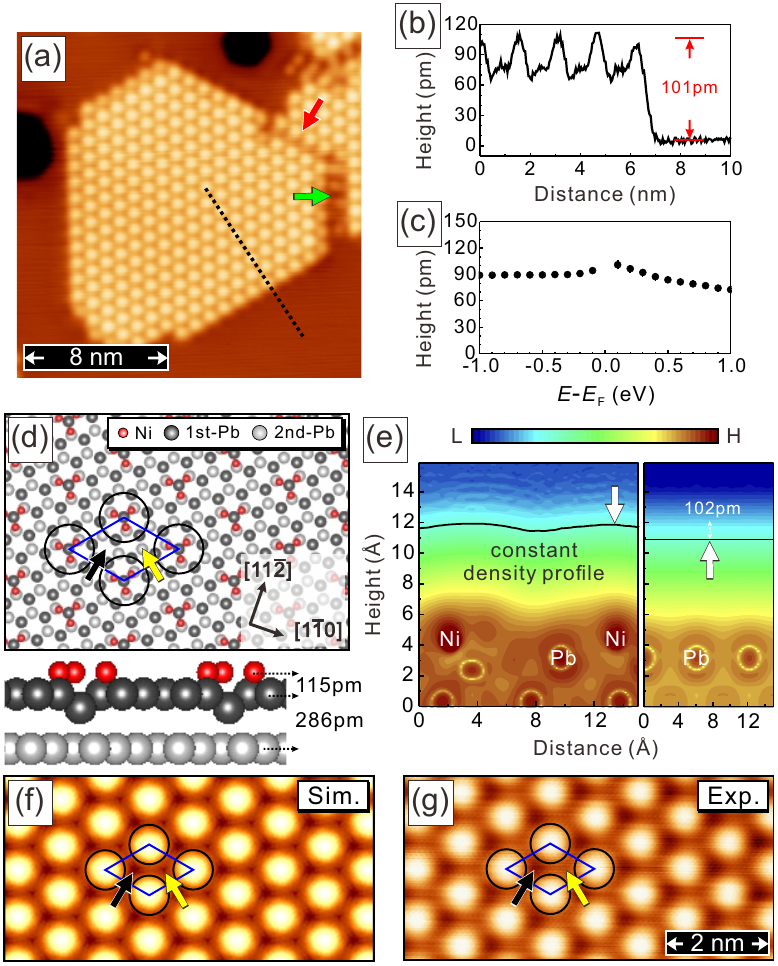}
\caption{(Color online)
(a)~Magnified STM image showing antiphase domain boundary (red arrow) and missing atom vacancies (green arrow) between neighboring $(\sqrt{7}\times\sqrt{7})R19.1^{\text{o}}$ Ni nanoislands ($U_{b} = +0.1$\,V, $I_{t} = 0.4$\,nA). (b)~The topographic line profile measured from the black dashed line in (a) represents an apparent height about 101 pm and bias dependence of apparent height has been shown in (c). (d)~Top panel: top view of structure model of $(\sqrt{7}\times\sqrt{7})R19.1^{\text{o}}$ Ni hexagonal lattice (blue rhombus) on Pb(111). Bottom panel: side view of structure model at which the adsorption height of Ni atoms is 115 pm and interlayer distance of Pb(111) is 286 pm. (e)~Surface charge density distributions of $(\sqrt{7}\times\sqrt{7})R19.1^{\text{o}}$ Ni on Pb(111) (left panel) and pristine Pb(111) (right panel). When tip and sample distance is about 6.3 \AA\,, the constant density profile displays an apparent height about 102 pm\, as indicated by two white arrows. (f)~Simulated STM image from the structure model in (d). The asymmetric height contrast inside the $(\sqrt{7}\times\sqrt{7})R19.1^{\text{o}}$ unit cell has been remarked by black and yellow arrows, which is in line with the experimental observation in (g) ($U_{b} = +0.1$\,V, $I_{t} = 0.4$\,nA).
\label{fig2} }
\end{figure}

Given high spatial and energy resolution, the tunneling conductance spectra have been carried out to access the proximity-induced superconductivity in $(\sqrt{7}\times\sqrt{7})R19.1^{\text{o}}$ Ni nanoislands grown on Pb(111). When the measurement temperature is below 7.2 K of the superconducting transition temperature ($T_{c}$) of Pb(111)\cite{JKim,JKim1}, the feature of superconducting gap ($\Delta$) opened at Fermi energy $E_{F}$ is therefore expected to appear in the $\mathrm{d}I/\mathrm{d}U$ curve acquired on the Pb(111) substrate, since it is the bulk material known for BCS-type electron-phonon mediated superconductor. According to the Fig. 3(a), the typical "U" shape of superconducting gap has been resolved on not only the Pb(111) (black curve), but also the $(\sqrt{7}\times\sqrt{7})R19.1^{\text{o}}$ Ni nanoislands. 

In order to have a quantitative comparison, the normalized $\mathrm{d}I/\mathrm{d}U$ spectra have been fitted to the BCS-like density of states (DOS) (red lines in the Fig. 3(a)), and the corresponding values of $\Delta_{Ni(\sqrt{7}\times\sqrt{7})} \approx 1.29$ meV\, and $\Delta_{Pb} \approx 1.25$ meV\, have been obtained. Interestingly, the $\Delta_{Ni(\sqrt{7}\times\sqrt{7})}$ is slightly larger than $\Delta_{Pb}$, which is counterintuitive to the typical understanding of proximity effect with a reduced size of $\Delta$ induced in the normal metal\cite{PGdeGennes,JJHauser,CWJBeenakker,CLambert}. We denote that the BCS fitting reasonably agrees with experimental $\mathrm{d}I/\mathrm{d}U$ spectra, especially in the critical energy range covering the superconducting gap, allowing us to extract the appropriate $\Delta$ values.

Besides the small enhancement of $\Delta_{Ni(\sqrt{7}\times\sqrt{7})}$, i.e., about 40 $\mu$eV\, as compared to $\Delta_{Pb}$, we have also found the proximitized superconductivity in $(\sqrt{7}\times\sqrt{7})R19.1^{\text{o}}$ Ni nanoislands is robust against the local perturbations of structural variation. As shown in the Fig. 3(b), the size and isotropic shape of $\Delta_{Ni(\sqrt{7}\times\sqrt{7})}$ as well as the coherence peak height do not exhibit a noticeable change according to nearly identical $\mathrm{d}I/\mathrm{d}U$ spectra measured on different atomic sites. This implies a transparent interface where the coherent Cooper pairs from the conventional \textit{s}-wave Pb(111) superconductor can penetrate easily to develop superconductivity in the atomic-thick $(\sqrt{7}\times\sqrt{7})R19.1^{\text{o}}$ Ni nanoislands.

Fig. 3(c) represents a series of temperature-dependent $\mathrm{d}I/\mathrm{d}U$ curves measured on the $(\sqrt{7}\times\sqrt{7})R19.1^{\text{o}}$ Ni nanoisland (blue) and Pb(111) (black). With an increase of temperature, both $\Delta_{Ni(\sqrt{7}\times\sqrt{7})}$ and $\Delta_{Pb}$ continue to decrease and their corresponding values as a function of temperature have been summarized in Fig. 3(d) (blue and black dots). They both vanish at about 7.2 K\, and thus yield to the same $T_{c}$, suggesting the superconducting $(\sqrt{7}\times\sqrt{7})R19.1^{\text{o}}$ Ni nanoislands are induced by the Pb(111) substrate via proximity effect. By using generic BCS gap equation\cite{JBardeen, RProzorov1}, not only the $\Delta_{Ni(\sqrt{7}\times\sqrt{7})}(T)$, but also the $\Delta_{Pb}(T)$ can be fitted reasonably well (blue and black lines in Fig. 3(d)). The fitting transition temperature $T_{c}$ is about 7.14 K\, in agreement with experimental results.


\begin{figure}[htb]
\includegraphics[width=\columnwidth]{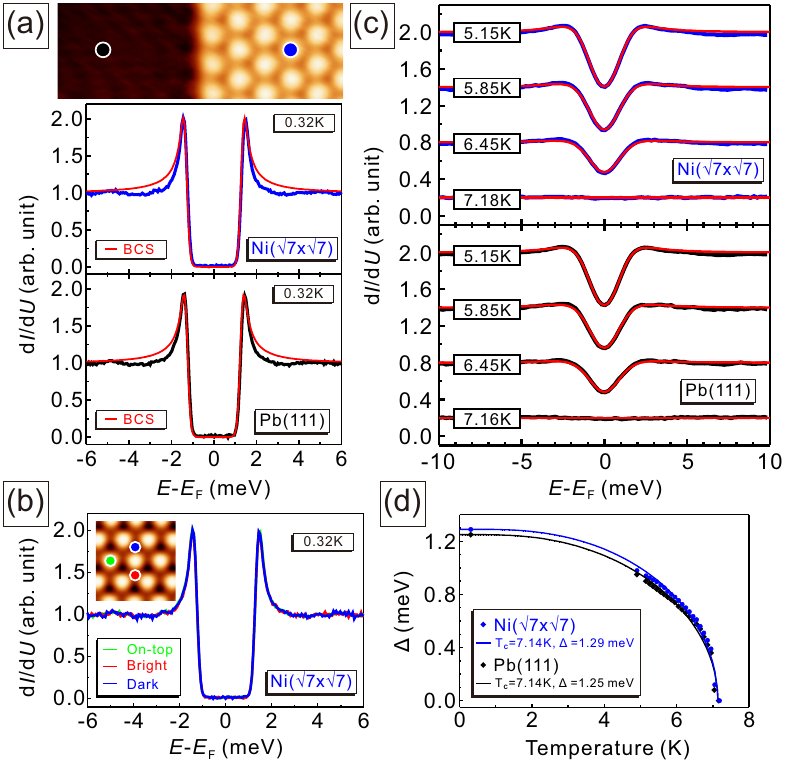}
\caption{(Color online)
(a)\, Point conductance spectra measured on the $(\sqrt{7}\times\sqrt{7})R19.1^{\text{o}}$ Ni nanoisland and the Pb(111) as indicated by blue and black points in topography at top. According to the BCS fittings (red lines), the $\Delta_{Ni(\sqrt{7}\times\sqrt{7})} \approx 1.29$ meV and $\Delta_{Pb} \approx 1.25$ meV have been obtained, respectively. (b)\, Point conductance spectra acquired on different atomic sites of the $(\sqrt{7}\times\sqrt{7})R19.1^{\text{o}}$ Ni nanoisland, including on top (green), bright bridge (red), and dark bridge (blue) as the topography shown in the inset. The proximity-induced superconducting state is robust due to the absence of considerable changes on resultant $\mathrm{d}I/\mathrm{d}U$ curves. (c)\, $\mathrm{d}I/\mathrm{d}U$ curves as a function of temperature on the $(\sqrt{7}\times\sqrt{7})R19.1^{\text{o}}$ Ni nanoisland (blue) and the Pb(111) (black). (d)\, Temperature dependent $\Delta_{Ni(\sqrt{7}\times\sqrt{7})}$ and $\Delta_{Pb}$ reveal the identical $T_{c}$ about 7.14 K as extracted from the universal BCS gap equation. (stabilization parameters: $U_{b} = +10$\,mV, $I_{t} = 1.0$\,nA for all $\mathrm{d}I/\mathrm{d}U$ curves)
\label{fig3}}
\end{figure}

In addition to the point conductance curve, the line spectroscopy taken point-by-point $\mathrm{d}I/\mathrm{d}U$ curve along the blue dashed line in the topography at top panel of Fig. 4(a) has been performed to map out the spatial dispersion of $\Delta_{Ni(\sqrt{7}\times\sqrt{7})}$ and $\Delta_{Pb}$ in the proximity region. The evolution of $\mathrm{d}I/\mathrm{d}U$ spectra as a function of spatial distance measured in the resolution of 4 \AA\, has been shown in the bottom panel of Fig. 4(a), representing the consistent isotropic shape of superconducting gap between the $\Delta_{Ni(\sqrt{7}\times\sqrt{7})}$ and the $\Delta_{Pb}$. Furthermore, the topographic line profile measured from blue dashed arrow line in the Fig. 4(a) has been quantitatively connected to the high energy resolution $\mathrm{d}I/\mathrm{d}U$ spectra in the Fig. 4(b). The spatial mapping of superconducting gap reveals not only the enhanced $\Delta_{Ni(\sqrt{7}\times\sqrt{7})}$, but also a sharp transition, i.e., within 1 nm\,, between $\Delta_{Ni(\sqrt{7}\times\sqrt{7})}$ and $\Delta_{Pb}$.


\begin{figure}[htb]
\includegraphics[width=\columnwidth]{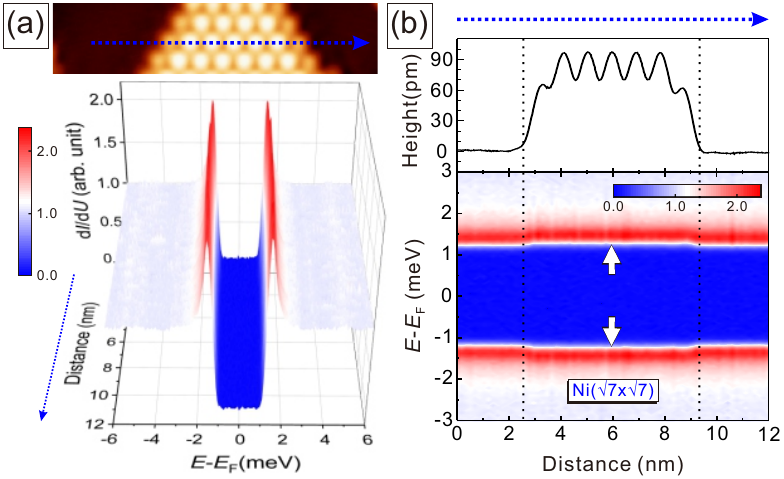}
\caption{(Color online)
(a) Spatial dependence of $\mathrm{d}I/\mathrm{d}U$ spectra measured along the blue dashed line in topography (top panel) represents the continuous evolution of $\Delta_{Pb}$ and $\Delta_{Ni(\sqrt{7}\times\sqrt{7})}$ in real space (bottom panel). (b) The topographic line profile (top) taken from the blue dashed line in topography of (a) has been compared to the spatial mapping (bottom) of $\Delta_{Pb}$ and $\Delta_{Ni(\sqrt{7}\times\sqrt{7})}$ directly, where the sharp transition from the discontinuity of superconducting gap as well as the small enhancement of $\Delta_{Ni(\sqrt{7}\times\sqrt{7})}$ (white arrows) have been clearly visualized. (stabilization parameters: $U_{b} = +10$\,mV, $I_{t} = 1.0$\,nA for all $\mathrm{d}I/\mathrm{d}U$ curves).
\label{fig4}}
\end{figure}

Apart from the small apparent height of $(\sqrt{7}\times\sqrt{7})R19.1^{\text{o}}$ Ni nanoislands grown on Pb(111), the DFT calculations further reveal that the net magnetic moment of Ni atoms vanishes could be explained by the charge transfer from underlying Pb atoms. On one hand, this vanishing magnetic moment leads to the absence of non-trivial in-gap states, e.g., Yu-Shiba-Rusinov (YSR) states\cite{AYazdani,SHJi,BWHeinrich,KJFranke} or zero-bias peak of MZM\cite{SNadjPerge,SJeon,MRuby,GCMenard}, because of the magnetism does not play a role in the superconducting $(\sqrt{7}\times\sqrt{7})R19.1^{\text{o}}$ Ni nanoislands. On the other hand, the electron acceptor of Ni atoms, which effectively causes hole-doping in Pb, provides a qualitative understanding for the observed superconducting gap enhancement in the STS measurements as discussed below. From the Bader charge analysis, each of the topmost layer Pb atom transfers about 0.1 electron to the Ni atom, which is equivalent to the hole-doping density of $n\approx0.942\times10^{14}cm^{-2}$ onto the Pb(111) substrate. To consider this hole-doping effect from the surface Ni atoms on the superconducting Pb(111) substrate, we have calculated the superconducting $T_c$ of bulk Pb as a function of the hole-doping level under the framework of the BCS theory. As shown in Fig.~5(a), the $T_c$ and $\Delta_0$ of Pb increases as the hole-doping level increases. They reach the maximum values of 7.46 K and 1.30 meV ($\Delta_0=2.032 k_B T_c$), respectively, at around 0.15~h/Pb, which corresponds to the hole charge density of $n\approx1.413\times 10^{14}cm^{-2}$. Then they decrease as the doping level raises further.
Figure 5(b) shows that the dome-shaped $T_c$ behavior is mainly governed by the dimensionless coupling parameter $\lambda = N_{F}V_{ep}$, while the other factors such as $N_F$, i.e., the DOS at $E_{F}$, and $\omega_{ln}$ play minor roles only.
The trend of the $V_{ep}$ strength in the $\lambda$ of Pb can be further elaborated by the $\alpha^2 F(\omega)$ spectra in Figures 5(c) and (d). For hole-doping concentration lower than $1.178\times10^{14}cm^{-2}$, the amplitude of the lower frequency peak of $\alpha^2 F(\omega)$ increases (red arrow in Fig. 5(c)) along with the increasing hole-doping concentration. Because of the denominator $\omega$ in the integration of $\lambda$ in the equation (3), the lower frequency part of the $\alpha^2 F(\omega)$ spectra contribute more strongly to the $\lambda$ and hence the $V_{ep}$ strength.
Consequently, the $T_c$ and the $\lambda$ increase for the lower hole-doping concentration, offering an explanation for the slight increase of $\Delta_{Ni(\sqrt{7}\times\sqrt{7})}$. Since there is about 3.2\% increase of $\Delta_{Ni(\sqrt{7}\times\sqrt{7})}$ as compared to $\Delta_{Pb}$ from the experimental $\mathrm{d}I/\mathrm{d}U$ spectra, this value is in line with the lower hole-doping concentration in Fig. 5(b) where the corresponding enhancement of $\lambda$ contributes to the enhanced $\Delta_0$ in Fig. 5(a). As for hole-doping concentration higher than $1.178\times10^{14}cm^{-2}$, not only the amplitude of the lower frequency peak is suppressed, but also this peak shifts toward the higher frequency significantly (blue arrow marked in Fig. 5(d)). As a result, the $T_c$ and the $\lambda$ decrease for the higher hole-doping concentration.

\begin{figure}[htb]
\includegraphics[width=\columnwidth]{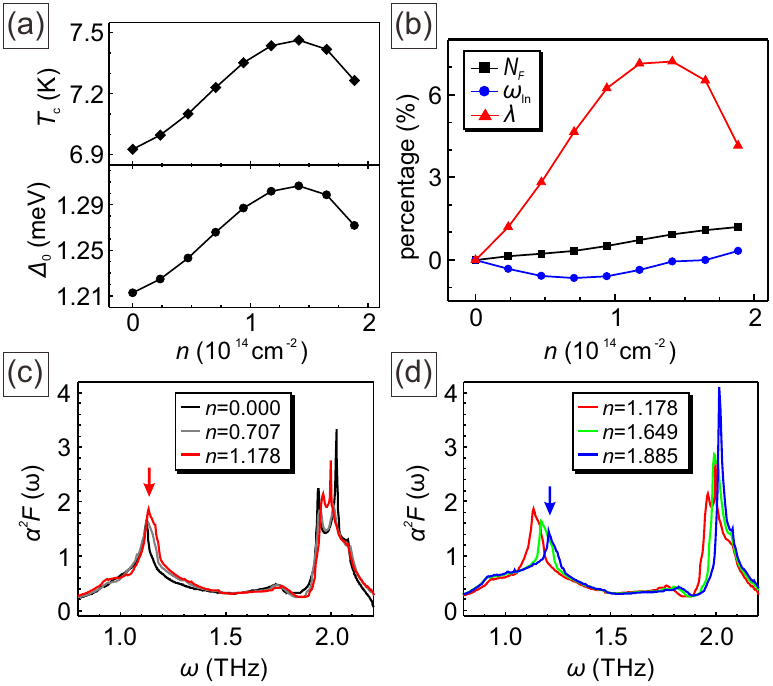}
\caption{(Color online)
(a) $T_c$ and $\Delta_0$ as functions of concentration of hole doping in bulk Pb. (b) Percentage change in terms of $N_F$, $\omega_{ln}$, and $\lambda$. The $\alpha^2 F(\omega)$ as function of hole-doping concentration for (c) smaller (d) larger than $1.178\times10^{14}cm^{-2}$.
\label{fig5}}
\end{figure}

\section{CONCLUSION}

In conclusion, we have investigated the proximity-induced superconductivity in the $(\sqrt{7}\times\sqrt{7})R19.1^{\text{o}}$ Ni nanoislands on Pb(111). By means of a low temperature growth at 80\,K, the Ni atoms energetically prefer to terminate at hcp site and develop the monolayer $(\sqrt{7}\times\sqrt{7})R19.1^{\text{o}}$ surface structure on the Pb(111). The charge transfer between the Ni and the Pb atoms results in the electron filling of 3\textit{d} orbitals leading to the vanishing magnetic moment of Ni atoms and a lack of magnetism in the $(\sqrt{7}\times\sqrt{7})R19.1^{\text{o}}$ Ni nanoislands. Furthermore, the STM simulations have been carried out to verify the asymmetric height contrast in atomic unit cell of $(\sqrt{7}\times\sqrt{7})R19.1^{\text{o}}$ lattice. Given high spatial and energy resolution, tunneling conductance ($\mathrm{d}I/\mathrm{d}U$) spectra have resolved the isotropic superconducting gap for both $(\sqrt{7}\times\sqrt{7})R19.1^{\text{o}}$ Ni nanoislands and Pb(111), and the $\Delta_{Ni(\sqrt{7}\times\sqrt{7})} \approx 1.29$ meV slightly larger than the $\Delta_{Pb} \approx 1.25$ meV has been extracted from the BCS fitting. On account of the same transition temperature $T_{c} \approx 7.14$ K, the temperature-dependent $\Delta_{Ni(\sqrt{7}\times\sqrt{7})}$ supports that the superconducting $(\sqrt{7}\times\sqrt{7})R19.1^{\text{o}}$ Ni nanoislands are proximity-induced from the bulk Pb(111) substrate. The small enhancement of $\Delta_{Ni(\sqrt{7}\times\sqrt{7})}$ has been further mapped out in real-space by the line spectroscopy, and the hole-doping effect from the surface Ni atoms could offer an explanation based on an increased $V_{ep}$ in fundamental BCS theory.




\section{Acknowledgments}

Y.H.L. and S.T. contributed equally to this work. D.S.L. and P.J.H. acknowledge support from the competitive research funding from National Tsing Hua University, Ministry of Science and Technology of Taiwan under Grants No. MOST-110-2636-M-007-006 and MOST-110-2124-M-A49-008-MY3, and center for quantum technology from the featured areas research center program within the framework of the higher education sprout project by the Ministry of Education (MOE) in Taiwan. J.H.T. acknowledges the financial support from the Ministry of Science and Technology of Taiwan under Grants No. MOST-109-2112-M-007 -034 -MY3 and also from the NCHC, CINC-NTU, and AS-iMATE-109-13, Taiwan.


\bibliography{DLFeIr111}
\bibliographystyle{apsrev}
\end{document}